\documentclass{aa}
\usepackage{graphicx,times}
 
\renewcommand{\arcsec}{\rm $^{\prime\prime}$ }

\begin{document}
   \title{\bf The RR Lyrae  Star Period -- K-band Luminosity Relation of the Globular Cluster M3}
\titlerunning{RR Lyrae period -- K-band Luminosity Relation in M3}

 \author{D. J. Butler}
  \offprints{D. Butler}
  \institute{Max-Planck-Institut f\"ur Astronomie, K\"onigstuhl 17, D-69117 Heidelberg, Germany \\
              \email{butler@mpia.de}
\thanks{Based on observations made with the NASA/ESA Hubble Space Telescope, obtained from the data archive at the Space Telescope Science Institute. STScI is operated by the Association of Universities for Research in Astronomy, Inc. under NASA contract NAS 5-26555.}
          }

   \date{Received 10 February 2003 / Accepted 24 April 2003}

   \abstract{
That the RR Lyrae star period -- K-band luminosity relation  
 is a promising tool as a distance indicator in
the Milky Way and Local Group of Galaxies is apparent on observational and theoretical
grounds in the literature.  Less clear is the 
 sensitivity of the relation, and consequently the physics of 
 horizontal branch stars, 
  to differences in stellar environment.
 In this paper, the first measurement of the (fundamental) period --
 K-band luminosity relation for
 the central region of  a globular cluster is presented. It is  based on
 a sample of seven RR Lyrae stars imaged with adaptive  optics. 
 In addition, the relation for the outer region has been reanalyzed, and is
 found to be in
 good agreement with both the previous estimate by Longmore et al. (1990), 
 and with the inner region relation, especially  when 
 irregular and double-mode RR Lyrae stars are excluded.
 Importantly, there is no difference between the slope of 
 the inner and outer region relation within measurement uncertainties,
 suggesting no difference in evolutionary state (luminosity). 
     Taking the M3 distance modulus as 15.0$\pm$0.07\,mag, the 
 period--absolute K-band magnitude relation derived by linear 
 least squares fitting is: \\
 ${M_{\rm K} = -0.96 \hspace{1mm}(\pm 0.10) - 2.42\hspace{1mm}(\pm
 0.16)\hspace{1mm} \rm Log\,P_{\rm o}}$ for the inner region.
 Excluding irregular variable stars, the  outer region relation is:
 ${M_{\rm K} = -1.07\,(\pm 0.10) - 2.38\hspace{1mm}(\pm 0.15)\hspace{1mm} \rm
 Log\,P_{\rm o}}$. This good agreement provides further  strong support 
 for the near-IR period-luminosity relation as a  distance indicator. 
   \keywords{Stars: variables: RR Lyr - Stars: distances - Stars: 
horizontal branch - Stars: imaging - Instrumentation: adaptive optics}
   }

   \maketitle
%

\section{Introduction}
For more than a century,    
  RR Lyrae stars, in addition to classical Cepheids, have played an important
 role as primary  distance indicators
  in  the Milky Way and the Local Group of galaxies. 
 The importance of RR Lyrae stars  stems primarily
  from their  brightness, being  presently
 observable  efficiently out to distances of several Mpc; they have 
 characteristic light curves,   easily measured
  periods, a narrow magnitude range, and a  high frequency in 
 many  globular clusters, 
 but their intrinsic luminosity 
is uncertain,  both empirically and theoretically (see Caputo et al. 2000,
  and reference therein).
 Indeed, the dependence of intrinsic luminosity
 on metal content is  uncertain:   metal-rich and -poor 
 RR Lyrae stars appear to 
 follow  different linear  M$_{\rm V}$ -- [Fe/H] relationships
 (Caputo et al. 2000).

In contrast, near-infrared observations of RR Lyrae stars 
 hold several distinct advantages
 over optical investigations:  they have  a significantly weaker  
  dependence on  [Fe/H] (Bono et al. 2001); 
 the  K-band amplitudes of RR Lyrae variables are smaller 
than their V-band values; and  their
 near-IR light curves are more symmetrical. All 
 these points lead toward more easily measured and  accurately determined 
 absolute magnitudes, and importantly, with a much reduced sensitivity to reddening. 
 The relationship between 
   the periods and K-band magnitudes of RR Lyrae stars has been measured
  in the pioneering semi-empirical 
 work of Longmore et al. (1990) who found a tight relationship for 
 several galactic globular clusters.
  Bono et al. (2001) provided the theoretical support for these findings.
 They calculated 
 the dependence  of the relation on metallicity, and 
 determined the  intrinsic dispersion in the three relevant parameters, namely 
period, luminosity and chemical composition. 
Confirming the reliability of the relation further, Bono et al. (2001)  
 applied the PL$_{\rm K}$ relation together with  predicted
  mass and luminosity from 
 current evolutionary models, to  
 forecast the  parallax of RR Lyr and found a good
  agreement with the HST result (Benedict 2002).

As quoted in the literature before, horizontal branch (HB) stars
 could be regarded has having crucial importance  in our 
 understanding of several unresolved astrophysical problems such as the
 second-parameter problem, the UV-upturn in elliptical galaxies (Ferguson
 1999), as well as the dependence of the absolute V-band magnitude 
 M$_{\rm V}$ of RR Lyrae stars  on chemical composition.

 A relevant key issue is abundance variations in stars. Abundance variations 
 (typically of C, N, and O) have been found to
 occur along the whole stellar evolutionary  sequence from  
 below the turn-off of the main sequence  to the red giant branch (RGB)   
  in several globular  clusters. It is hard to list all
 the significant reviews in this field, but  
 Kraft (1994), Cannon et al. (1998), Briley et al. (2001) and references
 therein provide an overview of current thoughts on the subject for giant
 stars,  while 
 on the topic of  main sequence stars, useful references can be found 
 in  Harbeck et al. (2003). 
   RR Lyrae star  luminosities, colours, and periods
  could be  affected  by 
  inhomogeneities in the initial
 chemical composition of the primordial gas, internal stellar mixing
  processes,  mass loss, and possibly 
 multiplicity in globular clusters,  and the cumulative effect of 
 such variations  may reveal themselves in precise 
 multi-wavelength period-luminosity relation studies.  
 It is fair to say that a  database of precise, consistent 
 measurements for a large number
 of RR Lyrae stars covering a range of environments (in globular clusters,  the
 field,  different galaxies, etc.) will be important for 
 useful future tests of the intrinsic dispersion predicted by 
 Bono et al. (2001).

In this paper, the period--K-band luminosity relation for the inner region  
 (r $<$ 20$^{\prime\prime}$) of the globular cluster M3 is presented. As a
 close  agreement with the outer region relation (r $>$ 50$^{\prime\prime}$) 
 is generally expected to exist but has not been shown, the 
  inner and outer region relations are compared.  
 The motivation has been to provide further support  
 for the near-IR period-luminosity relation as a 
 distance indicator.  This work has been made possible by  K-band
 adaptive optics -- assisted observations.

\section{Observations and data analysis}
\label{obs_proc}
\begin{figure} 
\centering
\includegraphics[height=8.5cm,width=8.5cm,angle=0]{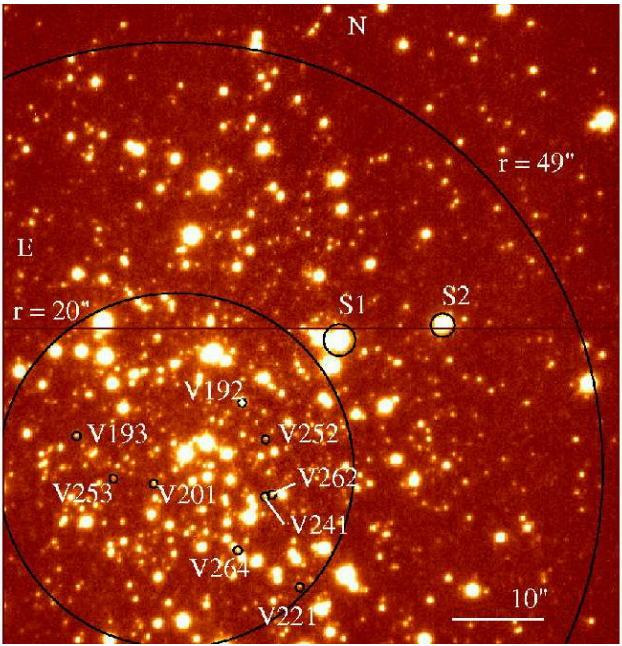}
\caption{Composite/cropped frame showing the central (K-band) field of M3 with
  the two AO guide stars, S1 and S2., and relevant RR Lyrae stars
  marked. North (N) and East (E) directions are included.}\label{fig1}
\end{figure}

The data consists of a series of K-band frames taken
 on April 19$^{\rm th}$, 21$^{\rm st}$  2000  
 using  natural guide star adaptive optics (AO) imaging with  
 ALFA (Kasper et al. 2000)  on the 3.5\,m telescope at the German-Spanish
Astronomical Centre on Calar Alto. The science camera was Omega Cass with a
 plate scale of 0.08\arcsec pixel$^{-1}$ giving a field of view (FoV)
 of 80\arcsec $\times$ 80$^{\prime\prime}$.  
  Dithering of source frames 
 during exposures was not performed, but there were occasional changes 
 of guide star between exposures.  Observing overheads 
  were of the order of a few $\times$10\%, and there were 
 target-of-opportunity override observations at the beginning of each night. 
 Conditions were photometric on the 19$^{\rm th}$ and 
  some light cloud was present on the 21$^{\rm st}$.
  The K-band source details  are given in 
 Table~\ref{table_april00_obs}.  On the
 21$^{\rm st}$  some J-, and H-band frames were  taken, but are 
 omitted as they are not within the scope of this study. 
  The Shack-hartmann 
 wavefront sensor (WFS) FoV  was set to 3\arcsec 
 to block out neighbouring
 stars, and a signal threshold level  was used to ignore faint stars in the WFS
 FoV.  There were two suitable AO guide (or reference) stars, namely  S1
 (m$_{\rm V}$ $\sim$ 12.8,
 m$_{\rm K}$ = 9.3) and S2  (m$_{\rm V}$ $\sim$ 13.2, m$_{\rm K}$ = 10.2), marked 
 in  Fig.~\ref{fig1}. 
 A consequence of the AO correction is that the best correction occurs
 in the direction of the guide star, becoming worse 
 at off-axis angles. There is no obvious evidence
 of systematic residuals due to radial  PSF changes in the 
 star-subtracted frames, though the central PSF flux is more pronounced
 toward the guide star. Through comparison with K-band  
 data from Davidge \& Grundahl, in preparation,  which is reported  
 in the next section, the effect on 
  photometric measurements is up to a few percent at most, 
 similar to reports by Davidge (2000) and references therein.
 The point spread function (PSF) FWHM varied between frames from about 0.3 to
 0.6\arcsec (see Table~\ref{table_april00_obs}) due to varying seeing, and 
 signal-to-noise.    The globular cluster M3 
 (NGC 5272; d$_{\rm galactocentric}$ = 12.2\,kpc; Harris 1996) was
 selected 
 because of its plentiful supply of RR Lyrae stars (Corwin et al., in
 preparation; Corwin et al. 2001; Clement 1997; Sawyer-Hogg 1973). 
 Several potential guide stars
 are within the central few arc-minutes, and there is a relatively low number 
 density of stars  compared to M15,
 which presently is the only other  cluster with both a  
 well studied,  and plentiful {\it central} RR Lyrae star  population (Butler et
 al. 1998; referred to hereafter as B98). 
  Additionally,  public Hubble Space Telescope (HST)
 integrations taken in 1995 April 25 are available and  were retrieved
 from the STScI science archive.  Calibration details are given 
 by Bagget et al. (2002)  and Holtzman et al. (1995).  The observations consist
 of two 70s  and two 3s  Wide Field Planetary Camera 2 (WFPC2) integrations
  with the F336W and F555W filters respectively (referred to here as U$_{\rm
 336}$ and V$_{\rm 555}$). They  are listed in Table~\ref{table_april00_obs}.
   The U$_{\rm 336}$
 data was used to  provide  accurate stellar astrometry, and both filters 
 were required for high  signal-to-noise giant star magnitudes.

\begin{table*}
\caption[]{Source information}
\label{table_april00_obs}
$$
         \begin{array}{llccllccc}
            \hline
            \noalign{\smallskip}
\rm  Date & \rm UT \hspace{1mm} Start & \rm Integration  & \rm Final^a    &
\rm  Date & \rm UT \hspace{1mm} Start            & \rm  Integration & \rm
Final^a \hspace{1mm}  & \rm Filter \\
      &      &            \rm  Time    &  \rm  Resolution             &   &   &  \rm   Time    &  \rm Resolution   & \rm    \\
     &      &           \rm   s        &  ^{\prime\prime}  &             &   &  \rm   s
        & ^{\prime\prime}   &     \\
\hline
 19/4/00  & 1:06:16   & 120 & 0.46 &  21/4/00 & 20:51:03    &  60 & 0.46& \rm K  \\
 19/4/00  & 1:09:03 &  60 & 0.45 &  21/4/00 & 20:59:35    &  60   & 0.46 & \rm K   \\
 19/4/00  & 1:11:38  &  60 & 0.44 &  21/4/00 &  21:13:57 &  60  & 0.40 &\rm  K     \\  
  19/4/00  & 1:20:58 &  60  & 0.32& 21/4/00 & 21:46:40 &  60  & 0.49 &\rm  K \\
 19/4/00  & 1:34:25 &  24  & 0.58 & 21/4/00 & 22:14:29  &  60  & 0.41 & \rm  K    \\
 19/4/00  & 1:46:24 &  60  & 0.42 & 22/4/00 & 0:20:40 &  42   & 0.44 & \rm K  \\
 19/4/00  & 1:48:35 &  60  & 0.40 &  22/4/00 & 0:28:12 &  63  & 0.44& \rm  K   \\
 19/4/00  & 1:59:22 &  60  & 0.38 & 22/4/00 & 0:39:24 &  39  & 0.54 &\rm  K   \\
 19/4/00  & 2:13:26 &  60  & 0.36 &  19/4/00 & 2:15:12  &  60   & 0.32  &\rm  K   \\
\hline
  25/04/95 &      -      & 2 \times 3 / 2 \times 70 & - & - & - &  - & - & \rm
     F555W/F316W  \\ 
\hline
             \noalign{\smallskip}
            \hline
        \end{array}
$$
\noindent $^a$ Measured at the centre of the PC1 region, about
20$^{\prime\prime}$ from the guide star(s); Uncertainty in spatial resolution
is $\sim$ $\pm$0.05$^{\prime\prime}$.
   \end{table*}

In the U$_{\rm 336}$ frame, the core of M3 is roughly centred on
the Planetary Camera (PC1) chip which has the highest angular resolution
 (0.045\arcsec pixel$^{-1}$). It was thus decided to concentrate on
 the PC1 frame only. The exact locations of stars have been obtained 
 by reducing the Planetary 
 Camera (PC1) frame following the procedure developed by B98. The resulting standard deviation (rms) of the astrometry residuals 
 from a high order U- to V-band coordinate transformation 
 are of the order of
 0.0025\arcsec in both 
 axes.  By using this astrometry  the 
 positions of stars in the K-band field have been accurately located. 
  For the present application, 
 the simple charge transfer efficiency ramp correction  
 of Holtzmann et al. (\cite{holtzman95a}) is adequate and has been applied.
 The U$_{\rm 336}$ and V$_{\rm 555}$ 
 have been zero point calibrated according to the Holtzman et al.
  (\cite{holtzman95a})  recipe
 (their Eq.\,(8) and Table 7).

For the near-infrared   data reduction  standard IRAF\footnote{IRAF is
  distributed by the National Astronomy Observatory, which is operated by the
  Associations of Universities for Research in Astronomy, Inc., under the
  cooperative agreement with National Science Foundation.
 } routines and scripts were used.  The
 basic image frames (typically 60\,s) were taken  
 with the cluster centre
positioned on one of the CCD quadrants,  and 
  a  standard 
 near-infrared  reduction scheme of  sky-subtraction and flat-fielding
 was followed.
 However, the measured sky frame illumination is
  non-uniform, varying  across each camera
 quadrant,  predominately along the readout direction.
 In contrast, the dome frames have a flatter appearance. 
 The  effect is such that the 
 drop-off in illumination intensity is up to 20\% in each quadrant. 
 The cause is not clear.
  For correction of this effect, the parameters involved
 are the number of counts per pixel from the  source
   S, the thermal signal  T, and the observed signal O which is 
 related to the normalized  flat-field response (or relative 
 quantum efficiency)  of the imaging system, k, by   
\begin{equation}
\rm O(x,y) = k(x,y) \times S + T(x, y)
\end{equation}
  The standard iterative procedure
 of fitting a low order surface function to the median filtered
 quadrants was applied to sky and dome flat-fields to determine the 
 conversion image
   k$_{\rm sky}$ /   k$_{\rm dome}$ (where S$_{\rm dome, sky}$ = 1.0).
  For each source image, the final flat-field was created 
 by taking the inverse-normalized lights `on' 
 and lights `off' dome frames  in turn, 
  then multiplying  each by the  appropriate conversion image, 
 and finally taking the difference between the two of them. 
 Importantly, this process
  removes the average bias and dark
 count in each pixel.

The next issue is sky subtraction. As K-band sky emission
  varies over time, sky frames taken close in time (less than 
 about two minutes) 
 to each object frame, which might otherwise swamp the light from the 
 faintest stars, need to be subtracted. 
 Sky frames were built by stacking image frames taken at different pointings
 at 300\arcsec from the cluster centre and dithered by 
 10-15$^{\prime\prime}$.  Incomplete  star removal in the bottom-right chip
 of some median sky frames  
 occurred due  to the presence of a few close 
 stars; but these stars are outside 
 the PC1 region and thus have no effect on the results presented in this 
 paper. 
   The best fitting function [k$\times$S]$_{\rm sky}$ (S$_{\rm sky}$ = 1.0) was determined for each stacked 
 sky frame; subtraction of it from the 
  sky frame itself yields the thermal component. 
 As it is important to remove most of the night sky flux, the sky 
 component (the best surface fit to the appropriate  sky frames)
 was scaled slightly for  most science frames, typically by about 1\%. 
 The thermal component, which is about 1-2\% of the sky flux  was then added.
 For hot/bad pixel removal, the flat-field  was multiplied by a hot pixel mask
 derived from  the flat-fielded sky frame. Next,
 for sky background removal,  the flux-weighted centroid of each
  source frame was determined and the frames 
 were shifted  to a common centroid. 
  This stacked source frame 
  and the  stacked sky frame, both of
  equal integration time, were then flat-fielded and  the difference was taken. 

 Following the calibration of all the K-band frames,  
  eleven frames  with sub- half arc second 
 spatial resolution were selected 
  for signal-to-noise/crowding reasons,  the 
 flux-weighted centroid of each frame was determined, and all were shifted 
 to a common centroid.  This was done three times using 
 a different  set of frames at each turn to allow 
 three  high signal-to-noise estimates of the K-band magnitudes.
 In this way,  a  bootstrap
 estimate of magnitude uncertainty can be obtained later
 after determining the photometry. To test
 the internal photometric precision later using the brightest stars
 the individual frames  were also reduced.

 It is possible that analysis of the non-PC1 portion of the full 
 K-band field in Fig.~\ref{fig1} may reveal some evidence  for red giant branch variable
 stars that could be considered in the context of giant star evolution;
  but such a study is outside the scope of this paper.

\subsection{Photometry}\label{photom}
The photometric reductions have been carried out using the DAOPHOT
  (Stetson 1987)  photometry package. 
 The procedure used  to derive U$_{\rm 336}$ and  V$_{\rm 555}$
  instrumental magnitudes, is taken  from that described 
 by  B98.
 The V$_{\rm 555}$ 
 astrometry is estimated to  be complete down to  V$_{\rm 555}$ = 17 (K $<$
  15 for (V-K)$_{\rm giants}$ $<$ 2) based on insertion of 
 artificial stars
  into the  V$_{\rm 555}$ frame and repeating the data  reduction. 

 PSF-fitting was performed
 using the routine ALLSTAR (Stetson \& Harris 1988),
 part of the DAOPHOT package.
  Each K-band frame has been considered as two  sections overlapping 
 by 20$^{\prime\prime}$, and 
   the following procedure for each section, taken partly from B98, has 
 been carried out:
  The initial PSF was obtained using standard recursive techniques;
  the star list at this point consisted only of stars detected by
 DAOFIND, which typically found about 250-300   in the PC1 region. 
 This allowed
 a high-order coordinate transformation 
 between the K-band frames and the HST/PC1 field, thereby accounting for 
 relative geometrical distortion near known RR Lyrae stars in the field.
   60-90 stars were selected  from each section. After applying this
 transformation, one for each section,  
 to the HST/PC1 star list, the rms positional accuracy was of the order of
 0.02\arcsec in both  axes.  Then,  a background-subtracted 
 image and its PSF were fed  into the 
 PLUCY (Hook et al. \cite{Hook94}) restoration task in IRAF. Although not 
  photometrically reliable, the data from the restoration task provided
 good initial magnitude estimates for the brightest stars 
 in the field. Together with these magnitudes, accurate coordinates,  
  sky background
   estimates and a unique ID number for each star, a final PSF fitting
 could be performed.  Next,   
 U$_{\rm 336}$\footnote{The light from the undetected stellar population 
  is dominated by light  from  upper main sequence stars in a narrow colour
 range; the deviation from the mean colour for the bulk of these 
  stars over the  field is expected to be up to a few tenths of a magnitude.} 
 background estimates were scaled up appropriately, followed by re-making 
 the PSF, and transformation of the list fitted by ALLSTAR using the 
 new star IDs, in  groups of 500 stars;  the sky background is 
 re-calculated
 for each star individually rather taking a group estimate
 as the background varies significantly across the field. 
 Finally,  the  photometry from each section was tied to
 the same internal zero point.   This step was performed by 
  taking a   calibration of  a  sequence of stars in the range
 K $<$ 17  against their counterparts in an average dataset.  Firstly,   
 the histogram of individual offsets,  typically forming 
 a normal-like distribution, was created. Then, in an iterative process, 
   histogram outliers  were ignored until  two different estimates of 
 the zero point  offset, the
 weighted-histogram average and median value,  differed
  by less than  0.01\,mag.
 In this way, 
  the calibration of the K-band magnitude zero point 
 is on a robust statistical footing, insensitive to large amplitude variable 
 stars and photometry outliers.  

In the context of photometric precision, it is worth mentioning that
 a  space-invariant empirical PSF made using 2 - 4 of the brightest stars,
 and modeled on  a moffat elliptical function of index 1.5 
 works best because the background light due to the faint unresolved stellar
 population  is distributed in  an inhomogeneous way.
 As there is a pointing difference of about 12\arcsec for some frames,
 it was possible to test the photometric precision along  both axes for
  each individual  frame; for this,   the way in which magnitudes differ from their time averaged/intensity weighted
 average was examined. No obvious indication of a trend with field
 position was found. 

\begin{figure} 
\includegraphics[height=9cm,width=4cm,angle=-90]{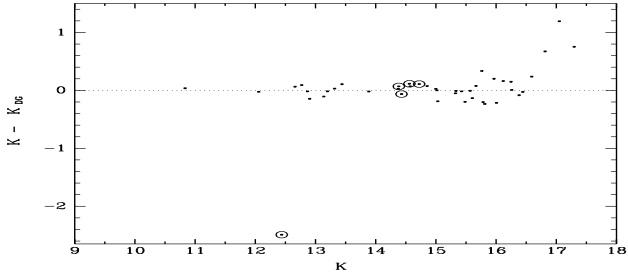}
\caption{Difference between the intensity weighted ensemble average K-band photometry 
 from this work and DG as a
  function of  K-band magnitude. The  RR Lyrae stars in common between the
  two data sets are marked with open circles. The circled outlier may be 
a  false measurement  due to stellar crowding. }\label{K_mag_residuals}
\end{figure}

\begin{figure} 
\includegraphics[height=9cm,width=5cm,angle=270]{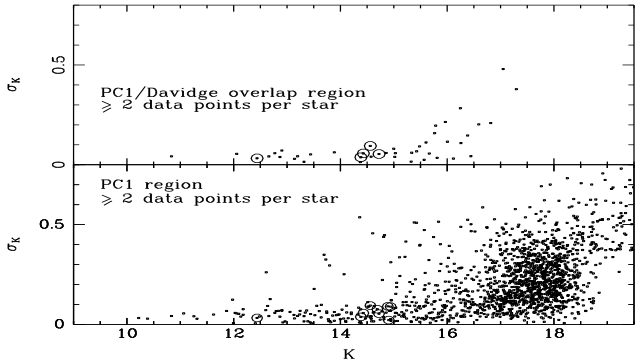}
 \caption{$\sigma_{\rm K}$ is plotted as  a function of mean K$_{\rm 
   PC1}$ magnitude. 
 {\it Top:}  Stars in common with the DG K-band data with at least
 two data points.
 The five known  RR Lyrae stars in common
 are marked with open  circles in both plots at K$\sim$15; 
 the circled point at about K=12.4 is an outlier in 
 Fig~\ref{K_mag_residuals}  and may be a false measurement due to 
 stellar crowding. {\it Bottom}: Stars in the PC1 region with at least 
 two data  points. Nine known variable stars (tabulated in
 Table~\ref{tableRRLy}) are marked with
 open  circles.}\label{sigmaK_versus_K}
\end{figure}

 For validation of the quality of the instrumental magnitudes  
 a  sequence of stars in the range
 K $<$ 17  was calibrated against their counterparts in  Canadian-France-Hawaii Telescope AO 
 data from Davidge \& 
 Grundahl (2002, in preparation; referred to here as DG).
 Davidge \& Courteau (1999) reported an absolute zero point error of
 $\pm$0.009\,mag for their K-band magnitudes  which came
 from CFHT AO observations; but conservatively, 
   $\sigma_{\rm K}$ = 0.05\,mag is assumed for the DG data. 
  The difference between the intensity
  weighted average K$_{\rm PC1}$  
 and the DG data  are compared in 
  Fig.~\ref{K_mag_residuals}; and  known 
  RR Lyrae stars from Corwin et al. (2002, in preparation) 
  are marked with open circles. Apart from 
 the outlying RR Lyrae star which 
 may be a false measurement due to stellar crowding, there is a good agreement.
  In Fig.~\ref{sigmaK_versus_K} the rms scatter is plotted as a function of
 K for stars in the  PC1 region with at least two (out of three) data points 
   (bottom); and   (top) for stars
  in common with the K$_{\rm DG}$ data  for
 which at least two K$_{\rm PC1}$ data points have been measured. 
 It can be seen that the rms value in the  RR Lyrae star 
 magnitude range is about 0.05\,mag,
   caused by both intrinsic
 variability and photometric error. It is useful to 
 note that intrinsic K-band amplitudes  
 of  RR Lyrae stars of type  RRab 
 are up to  a few tenths of a magnitude, and smaller for RRc-type variables.

\begin{figure} 
\includegraphics[height=9cm,width=7.0cm,angle=-90]{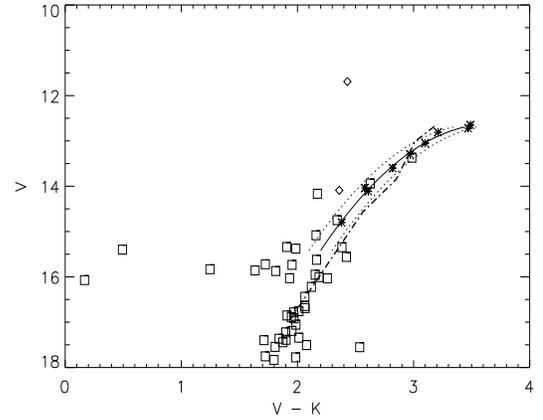}
\caption{K-band zero point offset assessment. 
 (V$_{\rm 555}$, V$_{\rm 555}$-K$_{\rm PC1}$)  CMD (squares) where
 K$_{\rm PC1}$ magnitudes have been shifted to match the lower end of the RGB  fiducial 
 ridgeline (dot-dashed/thick) deduced 
  from the work of Ferraro et al. (2002; referred to hereafter as F00) taking 
a distance modulus of 15.0. 
 The  Cohen
  et al. (1978; referred to hereafter as C78) giant branch data (asterisks and  diamonds)
  has been considered for the purpose of 
 zero point validation; taking the C78 data as comprising 
 asymptotic giant branch (AGB) stars at V $>$ 13.5, the 
 2nd order   polynomial/least squares fit 
  to the C78 data (excluding the outliers (diamonds)
 falls within $\pm$ 0.05\,mag of the  apparent AGB stars in the 
  K$_{\rm PC1}$ data. A $\pm$0.1 K-band magnitude envelope  (dotted)
 has been included  for convenience.
 See the final paragraph of Sect.~\ref{photom} for more information.}\label{CMD_zptcheck}
\end{figure}

For zero point validation, the  Cohen
  et al. (1978; referred to hereafter as C78) giant branch data 
  was considered.   The V-K colour of their M3  giant
 branch  ridgeline at M$_{\rm K}$ = -3.0 
 ($\sigma_{\rm V-K}$=$\pm$0.03\,mag) was checked
  and compared  with the value found on the  Ferraro et al. (2002; referred
  to hereafter as F00) ridgeline for globular clusters
 with a  similar metallicity to M3;  [Fe/H] =  - 1.34\, dex  from Carretta \&
  Gratton (1997) is adopted.  A good agreement was found
 to within  $\pm$0.05\,mag.  Next,  in order to 
  convert the F00 ridgeline  M$_{\rm K}$ to K-band magnitudes, 
  a distance modulus of 15.00$\pm$0.07\,mag (Caputo et al. 2000) 
   (see Sect.~\ref{harmony} later for a discussion) was assumed 
 and the V/K  data was overlaid with 
 the C78 V/K giant star data  on the (V, V-K)  colour
  magnitude diagram in Fig.~\ref{CMD_zptcheck}. A solid line fit
 to the C78 data (excluding the outliers (asterisks)) is included.
  The slight mismatch in
  ridgelines close to the  tip of the RGB arises because 
  the corresponding metallicity of the F00 line is only similar but not
 equal to that of M3. The lower end of the RGB is much less metal-sensitive,
  and so, as confirmed above, 
 it is safe to compare the ridgelines at two or more magnitudes
 fainter than the RGB Tip. Taking the C78 data as the reference
 dataset, the offset between the C78 and F00 
lines suggests that the lower end of the C78 ridgeline comprises
 asymptotic giant branch (AGB) stars at V $>$ 13.5.  Next,   the 
 (V$_{\rm 555}$, V$_{\rm 555}$-K$_{\rm PC1}$)  colour magnitude diagram 
 for stars with  $\sigma_{\rm K}$ $<$ 0.03\,mag and 
  $\sigma_{\rm V,\,555}$ $<$ 0.05\,mag (the 
 V-band zero point error is $\pm$0.05\,mag) was overlaid
 with the K$_{\rm PC1}$ magnitudes shifted by -0.1$\pm$0.05\,mag 
 to match the lower end of the  RGB fiducial ridgeline for [Fe/H]=-1.34\,dex, 
  (dot-dashed/thick) which was deduced from the F00 (V-K)--[Fe/H] relations.

\subsection{Deriving  $<$K$>$}
 The epochs of individual K-band frames are 
 distributed  over 20-50\% of RR Lyrae star light curve 
 cycles. Uncorrected,  the measured average
 magnitude  for each variable  might differ 
 from the intrinsic average value which is
 required for the PL$_{\rm K}$ relation.  
 This fact is also true for the deep/composite frames that are 
considered here for signal-to-noise reasons. 
 The adopted way to tackle
 this issue is taken from Jones et al. (1996; referred to 
 hereafter as J96): based on template near-IR light curves for both RRab- and
 RRc-type variables, 
  the method allows the offset from the true mean magnitude
 to be estimated if the period, optical amplitude, and 
 light curve phase of each exposure is known.  By taking frames
  from different light curve phases, 
  template shifting acts to minimize deviations from the 
 true mean magnitude arising from uncertainties in emphemeris phase (J96).
 For this reason, the magnitude offset associated with each composite  
 frames is determined by  calculating the offset associated with 
 each of its constituent frames, and then summing these values, 
 weighted by the constituent integration times. 
 The  magnitude offset associated
  with each star in any of the 
  deep frames  is in the range -0.05 to 0.03\,mag.
  The error reported for the J96 recipe is typically less than 0.03\,mag; 
 however, for each deep frame  there are eleven frames  covering a 
  range of light curve phases, leading to a negligible average error, and is
 ignored.

Zero-pointed photometry  for 
  a sample of seven RR Lyrae stars,  labelled  in  Fig.~\ref{fig1} according
 to Bakos et al. (2000),  was matched 
  to their associated periods (see Table~\ref{tableRRLy}). The RR Lyrae stars
  comprise five stars of  the type RRab and two stars 
 of the type RRc. 
   Period data  with a precision better than 1\% is taken from 
  CC01,  Corwin (2002, in preparation) and Strader et al. (2002).
 The epochs in Table~\ref{tableRRLy} are  
 taken from  CC02 which is based on their earlier work
 (CC01).  The CC01 study made use of  
 observations made on six nights in 1992, seven nights in 1993 
and one night in 1997. The observations for the present study were 
made in 2000. Of the  inner region variables considered 
 in this paper all have a precise emphemeris,
 (except V253 for which none has been reported).
  It turns out that
 the results in this paper are insensitive to a random choice of 
 emphemeris for this star,  as might be expected for an intrinsically low
  amplitude variable.  
 No  period change has been measured by CC01 for the inner region sample  
 considered here, although period change rates  
 up to about 0.5\,days Myr$^{-1}$
 were found for  38 outer region RR Lyrae stars.  Even if such changes
 were common to each star,    it is reasonable  
 to believe that the ephemerides that CC01  published are 
 valid for 2000. Incidentally, the  robustness of the PL$_{\rm K}$ technique 
 to period uncertainties is   further strengthened by 
 the template shifting mentioned earlier which  causes considerable
 reduction in  errors arising from errors in the light curve phases. 
  B-band amplitudes with a precision better than 5\%
 are available from CC01 for two of the 
 variables (A(B)$_{\rm V193}$= 1.6;  A(B)$_{\rm V201}$= 1.6) and 
 has been estimated for the remaining five variable stars
 using the A(B)--period plot of CC01.
 As noted later in
 Sect.~\ref{discussion}, adopting a conservative error of 
 $\pm$0.3\,mag for any of the B-band amplitudes,  
 produces a change in the measured slope
 of the period--luminosity relation of up to $\pm$0.01\,mag and 
 an even smaller error in the zero point of the relation.

Another issue is the possible presence of  Blazko variables
 in the inner region sample of RR Lyrae stars -- the light curve shapes
 of such Blazko-type stars could change on timescales of weeks or months.
  As none of the stars are labelled 
 as Blazhko-type variables in CC01 or CC02, there is 
 no `a priori' reason to suspect the presence of such variables 
 in the available sample but the possibility remains. 
  However, inspection of CC01's B-band light curves for Blazhko variables
 indicates that typical amplitude changes are  up to 
 about 0.5\,mag, in which case the derived PL$_{\rm K}$ would 
 be relatively insensitive to the effect.

  Next,  the intensity weighted average of the magnitudes
 determined from the three deep frames is considered.
  For stars with at least two valid
 measurements, the 
 adopted  recipe for intensity weighted averaging  of photometry
  is this:
    the magnitudes for each star are converted into a  flux
 level with a certain zero point, averaged and converted back to magnitudes. 
  The rms value was  calculated for each star; they vary in the  range 0.02 - 0.05\,mag.
 For line fitting later, 
 the instrumental magnitude error for each RR Lyrae star
 is taken as the average (arithmetic mean) of the 
  rms values for the seven variable stars,
  scaled down by $\sqrt{3}$, leading to  0.03\,mag. 
 In this way, statistical differences in magnitude errors 
 are smoothed out. 
 The rms value is included in the measurement of the slope of 
 a linear fit to the period versus K-band relation described in the 
 next section.

 The error in the absolute zero point of the K-band photometry
  from three deep frames is $\pm$$\sqrt{3}$$\times$0.01\,mag 
 ( relative zero pointing error); 
  plus the  error in the zero point offset is $\pm$0.05\,mag (systematic), leading
  to an absolute zero point error  of $\pm$ 0.053\,mag.

\subsection{Results}\label{results}
\subsubsection{Inner region}

 The fundamental period P$_{\rm o}$ has been calculated for 
 first-overtone (RRc) periods using the conversion  Log\,P$_{\rm o}$ = 
 Log\,(P$_{\rm RRc}$) + 0.127 (Iben 1974), 
 and is plotted against the calculated mean K-band magnitude  
  for each star in Fig.~\ref{slopes} (top-left).  
 Of the nine RR Lyrae stars  with  K-band magnitudes
  two stars V192, and V262 are outside the
 $\pm$ 1$\sigma$ envelope in Fig.~\ref{K_mag_residuals}
 as they are systematically too bright; only V262
 is plotted in Fig.~\ref{slopes} for the sake of presentation but 
  the measured mean magnitude and uncertainty
 are given for both in  Table~\ref{tableRRLy}. 
 V192 is    displaced to a  brighter magnitude than 
 the longest period RR Lyrae star;  CC01 classified the star
 as an RRab-type variable with P = 0.481d, 
 B = 14.94,  (B-V) = 0.554, and noted it as a 
  blend in their data.    For comparison,  
 RR Lyrae stars observed in M3  by CC01 
 have B-band magnitudes as bright as  $\sim$15.8.
  As  V192 is an outlier it is ignored in subsequent discussions.
 The second outlier is V262 which is
  marked with an open square in the left panel and has a small
  standard error (0.02\,mag). For the seven remaining RR Lyrae 
 stars there are three mean magnitude
 estimates, and the intensity weighted average is taken; one exception
 is V221  for which one measurement appears to be  spuriously 
 faint by 0.3\,mag  and the remaining two are used to 
 estimate the average. For the  
  measurement of the slope of the  PL$_{\rm K}$ relation, $\sigma_{\rm K}$,
   is taken to be  $\pm$0.03\,mag (rms internal scatter). 
 Both the data and the line fit 
 are shown in Fig.~\ref{slopes} (left-top). Coefficients of the 
 relation are tabulated in Table~\ref{table1_PLK}. 
 The  error in the true zero point of the relation 
 is made up of the error in the  absolute zero point ($\pm$0.053\,mag)
 of the K-band photometry  plus (in quadrature) the error in
  the PL$_{\rm K}$ zero point calculated by the line fitting
 ($\pm$0.05\,mag)  which leads to 
 $\pm$0.073\,mag.  An additional error must be included
 for the  M$_{\rm K}$--period relation. 
 For reasons explained in Sect.~\ref{harmony}, the adopted distance modulus
 is 15.07$\pm$0.07\,mag, which leads to a zero-point error of 
 $\sigma_{\rm M_{\rm K},\,zero\,pt.}$=$\pm$0.10\,mag in
 the relation. For different future distance modulii  
 this  uncertainty must be changed accordingly.

 Out of academic interest, the relation for
 the photometrically superior DG data set (FWHM $\sim$ 0.14$^{\prime\prime}$) is 
 plotted in Fig.~\ref{slopes} (top-right), shifted by -0.1\,mag$\pm$0.05\,mag
  to match the  K$_{\rm PC1}$ -band photometry (zero pointed to C78); conservatively, 
  the uncertainty in  the DG photometry is taken to be 0.05\,mag for the line fitting.
The outlier marked with an open square at K=15.87
 is V215 in  CC01 and has been ignored during line fitting. 
 There is a significant difference 
 between the best fit line in this panel
 and that of the outer region relation (bottom), most  likely because
  no J96 correction has been applied as the observing epoch is absent.

\subsubsection{Outer region}\label{PLK}
 The  work by Corwin \& Carney (2001; referred to here as CC01)
 provides  up to date information on  
  RR Lyrae star periods and classification for the outer
  cluster  region.  The Longmore et al. (1990; hereafter referred
 to as L90) data was  reanalyzed using the new
 periods for the cases of inclusion and exclusion
 of   stars classified as 
 Blazhko or double-mode (RRd) RR Lyrae stars. 
 In this way, the sensitivity of the slope to these stars can 
 be estimated, in addition to improving the precision of the relation.
  K-band magnitudes were taken from L90 which have an  error
   $\sigma_{\rm K}$ = $\pm$0.06\,mag,  and the latest period values
  come from  CC01. 
 There are 48 stars with K-band magnitudes from the 
  L90 study,  and they cover the radial range 2.6-13.3\,arc minutes. 
 Of these stars,  37 are now known to be 
 of type RRab, nine are  of type RRc, one is of type RRd and one,
  V113,  is not in the CC01 catalogue of variable stars in M3.  However,
  V113 has been  recorded by Szeidl (1965) as an RR Lyrae star and is 
 therefore included. The period of V113 used by L90 is 0.513d;    
 the same  value is recorded in the Clement (1997) catalogue of variable stars.
 V113 is not classified
 by  Clement (1997)  but is fairly safe to take it as an 
  RRab-type variable based on its period.
 The RRd star, and 15 of the RRab stars which are classified as  Blazhko
 (irregular) RR Lyrae stars by CC01 have been ignored for the 
 fitting; they are marked by open circles in Fig.~\ref{slopes} (bottom). 
 The gradient and zero point of the PL$_{\rm K}$ relation are 
 tabulated in Table~\ref{table1_PLK} for the cases of 
  inclusion and exclusion  of irregular RR Lyrae stars (rows 2 \& 3 respectively).
 The error in the zero point of the absolute PL$_{\rm K}$
 relation consists of $\pm$ 0.06\,mag (internal zero point error); 
 $\pm$ 0.04\,mag (rms fitting error); and   $\pm$ 0.07\,mag (distance modulus
 uncertainty -- referred to in the previous section). This leads to 
 a zero point error of $\pm$0.10\,mag.

\section{Discussion}\label{discussion}

\subsection{Harmony in the PL$_{\rm K}$ relation}\label{harmony}

Together with accurate period data from the literature, the 
 K-band observations prove on good photometric grounds that 
  there is  no apparent difference between the inner
 and outer region  PL$_{\rm K}$ relation. This conclusion is based on the following: 

 \noindent 1.  The gradient 
 of the inner region relation is weakly
 dependent on the way $<$K$>$ is calculated (intensity- or 
  magnitude-weighted  average) as shown in Table~\ref{table1_PLK}. 
 The agreement is dependent  on the J96 magnitude correction for RRab-type stars;
 excluding the correction produces a significantly steeper slope.
  The agreement between the inner and outer region relation 
 is not dependent on uncertainty in A(B); 
 an error of  $\pm$30\% in any of the amplitudes 
  produces a change in slope
 of $\pm$0.01, which is negligible  compared to fitting errors,
  and an even smaller error 
 in the zero point.  For both the inner and outer region relations, 
 periods from CC01/CC02 (and Clement 1997 in the case of V113) 
 are well determined to a precision 
 significantly better than 1\%. CC01 mean B-band magnitudes 
 are precise to at least 5\%.

\noindent 2. The absolute K-band extinction, A$_{\rm K}$, 
 can be derived using A$_{\rm K}$/E(B-V) =  0.13 
  (Cardelli et al., 
 1989). As E(B-V) = 0.01 (Dutra \& Bica, 2000) which  is negligible,  
  no correction has been made for K-band atmospheric extinction 
 which would be of the order of 0.001\,mag.

From these findings I conclude the following:

\noindent 1.  Including the irregular variables, but taking 
 improved periods, there is good agreement 
 between the L90 slope measurement -2.35$\pm$0.15 for the outer region
  and value  from an updated
 analysis of the L90 data  (-2.35$\pm$0.14). 
 There is also a fine agreement with the 
 ensemble average of the relation -2.338$\pm$0.067 determined by 
 Frolov \& Samus (1998) who reanalyzed the period/K-band magnitude 
 data tabulated by L90 for nine galactic globular clusters.

\noindent 2.  Inclusion or exclusion of the irregular variable stars has marginal 
 impact on the gradient and zero point of the relation (see
 Table~\ref{table1_PLK}, rows 2 \& 3); this arises because the  irregular
 stars are well-distributed about the PL$_{\rm K}$ line.

\noindent 3.  There is good agreement between the PL$_{\rm K}$ relation 
 for the inner and outer regions (see Table~\ref{table1_PLK}, rows 1, 2, \& 3).

\noindent 4.  For the absolute K-band magnitude -- period relation,
 the distance modulus to taken to be 15.0$\pm$0.07\,mag (Caputo et al. 2000,
 hereafter referred to as C00)  and the  reasoning behind the value chosen is explained as follows:
  It is true that the distance modulus for M3 is uncertain with values in the 
 range  14.8-15.2 appearing in the literature (C00). 
 Setting observational errors aside  for the sake of clarity, 
 the uncertainty arises because
 in order to measure a distance modulus in a certain bandpass, 
 the average  magnitude of a sample of RR Lyrae stars 
 in that bandpass, corrected for reddening, is measured through observations 
 while the absolute magnitude is estimated 
 through modeling.
 Thus, as a  metallicity and a luminosity must be assumed for modeling, 
 affecting the model stars' magnitude differently in different bandpasses,
  there may be  a mismatch between the 
 distance modulus derived for one bandpass and  
  the value derived for another bandpass. 
 However, Bono et al. (2001)
 report a way to find an average  luminosity
  for RR Lyrae stars by determining the luminosity 
 that gives the same visual  and IR distance modulus,
 thereby yielding a mean distance modulus. For M3 they find (m-M)$_{\rm K}$ =
 15.03$\pm$0.07\,mag which they note as being in good agreement with the 
  mean V-band distance modulus of 15.00$\pm$0.07\,mag derived by C00.
 As Bono et al. (2001) concluded for this point,
  this harmony proves the internal consistency 
 of the pulsation modeling and reinforces the belief in the literature that 
 the reddening towards M3 is almost zero.
Adopting the  C00 distance modulus  for the 
  inner region relation, 
 the absolute K-band magnitude for the inner region  is:

\begin{equation}\label{MKeqn}
\rm M_{\rm K} = -0.96\pm 0.10 - 2.42\pm0.16\,\rm Log\,P_{\rm o} 
\end{equation}

There is good agreement with 
 the slope of the empirical metal-free relation available in the literature
 provided by Jones et al. (1992),

\begin{equation}\label{MKeqn1}
\rm M_{\rm K} = -0.88\pm 0.06 - 2.33\pm0.20\,\rm Log\,P_{\rm o} 
\end{equation}
 
 which  differs significantly from the slope of 

\begin{equation}\label{MKeqn2}
\rm M_{\rm K} = -1.07\pm 0.10 - 2.95\pm0.10\,\rm Log\,P_{\rm o} 
\end{equation}

by Skillen et al. (1993). For the outer region, excluding 
 irregular variables, one has 

\begin{equation}\label{MKeqn3}
\rm M_{\rm K} = -1.07\pm 0.10 - 2.38\pm0.15\,\rm Log\,P_{\rm o}. 
\end{equation}

\subsection{Comparison with theory}
Bono et al. (2001) have recently calculated the theoretical PL$_{\rm K}$
relation using non-linear convective stellar pulsation   
  models over the range of stellar temperatures covered by the 
 instability region.  
  They considered a  range of  masses,  chemical composition, and luminosities,
  and then assessed the sensitivity of the relation to  these parameters.
 They derived that absolute K-band magnitude
 of RR Lyrae stars with metal content in the range 0.0001 $<$ Z $<$ 0.006 
 is correlated with period and metallicity as

\begin{equation}\label{eqnPLZ}
\rm M_K =0.139 - 2.07 (\rm Log\,P_{\rm o} + 0.30) + 0.167\,Log\,Z 
\end{equation}

with a total intrinsic dispersion of $\sigma_{\rm K}$ = 0.037\,mag, including
 a luminosity uncertainty of $\Delta$\,Log\,L $\sim$
 $\pm$0.04, and mass variations of 4\%.

 The empirical study of L90 found good agreement between the PL$_{\rm K}$ 
 relations measured for several galactic globular clusters which suggests that
 both the evolutionary (i.e. luminosity) effects and the spread of
 stellar masses inside the instability strip marginally affect that
 relation. On the theoretical side, Bono et al. (2001) 
  concluded that  for a certain metallicity
 the predicted PL$_{\rm K}$ relation  is marginally dependent on stellar mass 
  uncertainties. A  mass difference 
  would appear as a zero point offset in the PL$_{\rm K}$ relation. 
  In this context, it is worth remarking that  
 CC01 tentatively suggested a possible 
 scenario to explain the presence of some 
 optically sub-luminous RR Lyrae stars observed
   toward the central region of 
 M3, namely that     collisions due to stellar crowding
 might  prematurely halt 
 core  helium burning, leading to a lower luminosity than normal.

\begin{figure*} 
 \centering
\includegraphics[height=14cm,width=15cm,angle=-90]{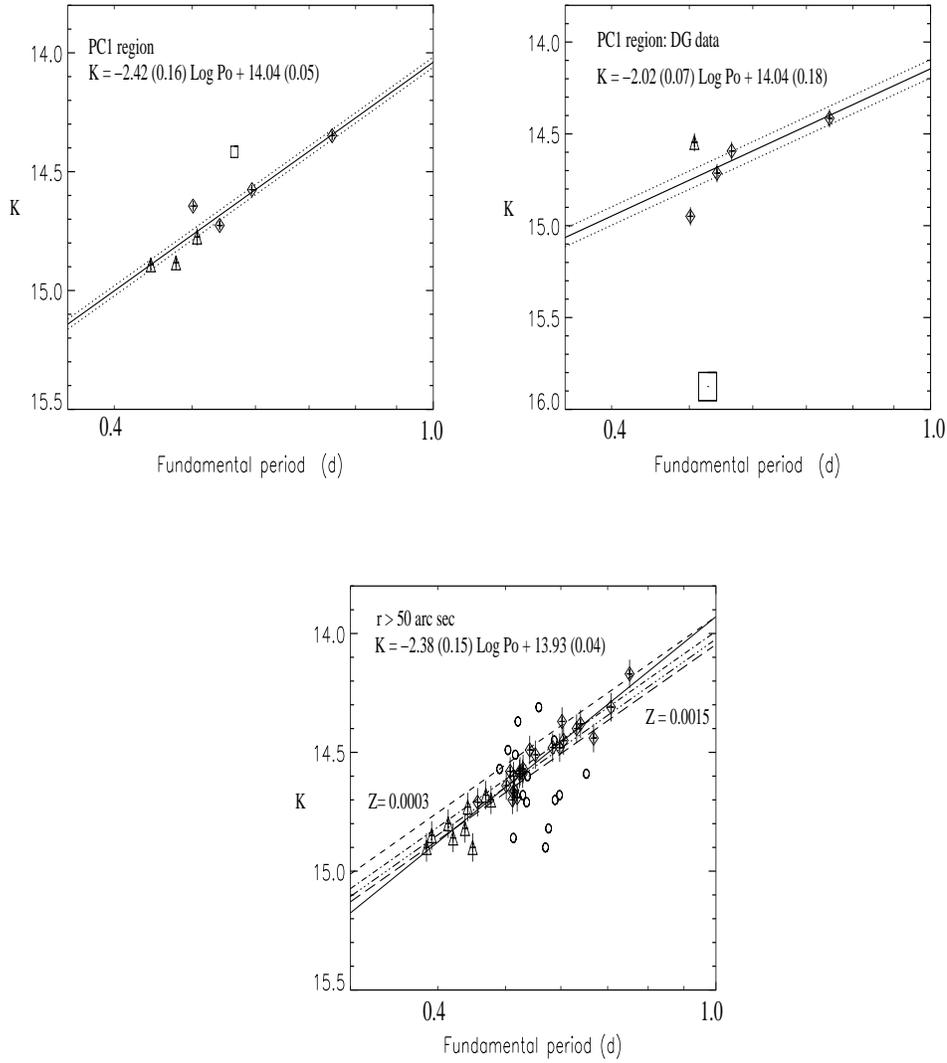}
\caption{Linear least squares fit to the  K-band magnitude / period data (solid). The
   dotted lines form the 1$\sigma$   envelope  of the linear least squares fit
 to the data.    Coefficient errors of the general relation a\,Log\,P$_{\rm
   o}$ + b, arising from the line fitting are given in parentheses. {\it Top-left:} 
 Diamonds and triangles mark the  RRab- and
   RRc-type RR Lyrae stars respectively.  
{\it Top-right:} DG data shifted by -0.1\,mag to match the K$_{\rm PC1}$ data
 (zero pointed to match C78), with no J96 offset applied 
  as the epoch is unavailable. The open square marks an outlying 
 star; it is  V215 (K=15.87\,mag
 in the DG data and is not in the  K$_{\rm PC1}$ data). 
 {\it  Bottom:}    Outer region data. 
  Z =  0.0003, 0.0007, 0.0011 and 0.0015  tracks are overlaid (from upper to
 lower tracks). 
 Data points marked with open circles are the Blazhko (irregular) stars  
 and one RRd-type  RR Lyrae star. 
 See text in Sect.~\ref{results} for further information.}\label{slopes}
\end{figure*}

Another relevant issue is metallicity.  Reviewing  the literature,
one has

\begin{equation}\label{eqn2}
\rm M_K = -0.647 - 1.72\, Log\,P_o  + 0.04\,[Fe/H]
\end{equation}

\begin{equation}\label{eqn3}
\rm M_K = -0.76 - 2.257\, Log\,P_o   + 0.08\,[Fe/H]
\end{equation}

and 
\begin{equation}\label{eqn4}
\rm M_K = -0.72\pm0.11 - 2.03\pm0.11\,Log\,P_o   + 0.06\pm0.08\,[Fe/H]
\end{equation}

by L90, Liu \& Janes (1990), and Jones et al. (1992) respectively. 
 Setting  [Fe/H] = -1.34\,dex (Carretta \& Gratton 1997) in  Eq.\,(~\ref{eqn2}) and (~\ref{eqn3}) gives M$_{\rm
 K}$ = -0.18 and  -0.19\,mag respectively  for  Log\,P$_{\rm o}$ =  -0.3.
 Eq.\,(~\ref{eqn4}) gives   a similar result.
In contrast,  the IR flux method that gives Eq.\,(~\ref{MKeqn3})
  yields  M$_{\rm K}$ = -0.36, which is
 significanty brighter; this is the same discrepancy noted by
  L90 and Bono et al. (2001).
  Next,   adopting a distance modulus of 15.0\,mag, tracks of constant Z were
 derived from     Eq.\,(~\ref{eqnPLZ}) and overlaid in  Fig.~\ref{slopes} (bottom). Measurement errors, especially in the 
 zero point  forbid a useful fit, but there is  fair qualitative agreement 
 between the data and  log Z = -3.04\,dex which corresponds to 
   the adopted M3  metal abundance  [Fe/H] = -1.34\,dex.  It is 
 hard to draw a more substantial conclusion from the graph 
 due to L90 measurement uncertainties ($\pm$0.06\,mag)
 and the distance modulus uncertainty ($\pm$0.07\,mag).  
 Solving  Eq.\,(~\ref{eqnPLZ}) yields M$_{\rm K}$ = -0.367\,mag
 in fine agreement with the IR flux method (Eq.\,(~\ref{MKeqn3})).
  Hence, the   PL$_{\rm K}$ relation, a metal-free relation, is in good
agreement with the predicted 
PLZ$_{\rm K}$ relation, meaning that the observations are fairly well matched
 by the results of stellar evolutionary models.
 This study cannot address the issue of intrinsic dispersion in the 
 PL$_{\rm K}$  relation for signal-to-noise reasons.
   Incidentally, the signal-to-noise ratio of stars  in the central 
 regions of globular clusters is affected by  stellar  crowding quite severely.
 To explore the physics of stellar evolution and pulsation theory
 through the PL$_{\rm K}$ relation in the central regions of 
 globular clusters, very high angular resolution imaging is required, and is  becoming  
 available with adaptive optics assisted near-IR imagers, such 
 as NAOS-CONICA on the ESO VLT,  
 and later  with  NIRCam on the James Webb Space Telescope.

 \begin{table*}
\begin{tabular}{llllllll}
            \hline
 \rm ID          & \rm  \rm Epoch                 & \rm Period   & \rm Type
 & $<$K$>$  & \rm A(K) & A(B) \\ 
    &    \rm  (JD)         & \rm (days)   &  & (mag) & (mag) & (mag) \\
  (1)  &   (2)          &   (3) & (4) & (5) & (6) & (7) \\
\hline
v192 & 2449090.225   & 0.49790  & ?  & 12.44$^{\rm b}$  & - & -  \\   
v193 & 2449090.327  & 0.74784  & RRab &  14.35   & 0.34$\pm$0.01 & 1.6$^d$$\pm$0.1  \\  
v201 & 2449090.133  &  0.59408  & RRab &  14.73    & 0.34$\pm$0.01  &
1.6$^d$$\pm$0.1     \\  
v221 & 2449090.125  & 0.378752 & RRc &  14.77  & 0.27$\pm$0.03  & 0.75$\pm$ 0.3\\ 
v241  & 2449090.014  &  0.59408 & RRab &  14.58  & 0.28$\pm$0.03  & 1.0 $\pm$ 0.3 \\ 
v252  &  2449090.157  & 0.50155 & RRab  & 14.64   & 0.33$\pm$0.03 & 1.5 $\pm$ 0.3\\
v253 & -   &  0.33161 &  RRc &  14.89  & 0.27$\pm$0.03  &  0.75$\pm$ 0.3 \\ 
v262 & -    &   0.5647$^{\rm a}$  & -  & 14.41$^{\rm c}$  & 0.29$\pm$0.03   & 1.1 $\pm$ 0.3\\
v264  & 2448755.093  & 0.35649 & RRc & 14.88  & 0.27$\pm$0.03  & 0.75$\pm$ 0.3\\ 
\\
 \hline
\end{tabular}
\noindent 
\\
Inner region variable stars. \\
Cols. (1) ID. V241 and V252 are X17 and KG4 respectively in CC01/CC02;
  (2) Julian date from CC02;
 (3) Periods. All  from   CC02  except $^{\rm a}$  Strader et al. (2002);
(4) RR Lyrae type from CC02; 
(5) Intensity-weighted K-band magnitude, $\sigma_{\rm K}$ = $\pm$0.03\,mag
 (random), plus $\pm$0.05 (systematic) or $\pm$0.058\,mag combined; 
 $^{\rm b}$ ignored as it is an outlier. No J96
 correction applied as the magnitude is inconsistent with an RR Lyrae star;
$^{\rm c}$ is an ignored outlier in Fig.~\ref{slopes} (top-left).
 (6) Expected K-band amplitudes determined by applying Eq.\,(7) of J96.;
  (7)  $^{\rm d}$  from CC01. A conservative error has been adopted.
  Other A(B) values  
 were taken with conservative errors from the CC01 
 A(B) - period graph, their Fig. 9.
     \caption[]{Relevant details for  nine variable stars from CC02
  in the K$_{\rm PC1}$ 
 field.}\label{tableRRLy}
\end{table*}

   \begin{table*}
  \begin{tabular}{l|ll|ll|l}
          \hline 
  \rm  Region     &  \rm  Gradient$_{\rm int}$ & \rm  Zero point$_{\rm int}$ &
  \rm  Gradient$_{\rm mag}$ & \rm  Zero point$_{\rm mag}$ & Comment \\
  \rm       &  \rm  a & \rm  b & \rm        a & \rm  b  \\
    \hline 
  \rm  Inner     &  \rm  -2.42 (0.16)  & 14.04 (0.05) & -2.40 (0.17) & 14.04
  (0.05) & this study \\
  \rm  Outer     &   - & - & \rm  -2.38  (0.15) & 13.93   (0.04)   &  new analysis
  of L90 data, excluding Blazhko variables \\
  \rm  Outer     &   - & - & \rm  -2.35  (0.14) & 13.95 (0.04)   &  new analysis
  of L90 data,
  including Blazhko variables \\
    \rm  Outer      &   - & - & \rm  -2.35 (0.15) & 13.95 (0.04) & L90 included Blazhko variables  \\
    \hline 
\end{tabular}
     \caption[]{Coefficients of the general relation 
 $<$K$>$ =  a Log P$_{\rm o}$ +   b. Values in parentheses are the
 uncertainties; int and mag indicate where intensity- and magnitude-weighted
 average magnitudes respectively have been used.}
\label{table1_PLK}
\end{table*}

\begin{acknowledgements}
 The support of S. Hippler, M. Kasper, and the 
 rest of the ALFA-team during observations is much appreciated.
  T. Davidge is thanked for providing 
 his K-band magnitude data, as is  M. Corwin for the updated
 list of RR Lyrae stars  with their  periods. 
 The author acknowledges the support of the research and training network on
 `Adaptive Optics for Extremely Large Telescopes' under contract
  HPRN-CT-2000-00147. 
 It is a pleasure to thank the
  anonymous referee  for several very helpful comments and suggestions.
\end{acknowledgements}

\end{document}